\documentclass[conference]{IEEEtran}

\usepackage[normalem]{ulem}

\usepackage{mathrsfs}
\usepackage{amsmath}

\usepackage{graphics}
\usepackage{listings}
\usepackage{xspace}
\usepackage{amsmath}
\usepackage{algorithm}
\usepackage{multirow}
\usepackage{rotating}
\usepackage{hhline}
\usepackage{blindtext}
\usepackage[noend]{algpseudocode}
\usepackage{etoolbox}%
\usepackage{enumitem}
\usepackage[colorinlistoftodos]{todonotes}
\usepackage{xcolor}
\usepackage[most]{tcolorbox}
\usepackage{verbatim}
\usepackage{courier}
\usepackage{amsmath}
\usepackage{listings}
\usepackage{enumitem}
\usepackage{soul}

\definecolor{mygreen}{rgb}{0,0.6,0}
\usepackage[font=small,labelfont=bf]{caption}
\usepackage{setspace}

\newcommand{\overhead}{45}

\begin{document}







%

\title{MOARD: Modeling Application Resilience to Transient Faults on Data Objects}

%

\author{
\IEEEauthorblockN{Luanzheng Guo}
\IEEEauthorblockA{EECS, University of California, Merced \\
lguo4@ucmerced.edu}
\and
\IEEEauthorblockN{Dong Li}
\IEEEauthorblockA{EECS, University of California, Merced \\
dli35@ucmerced.edu}
}

\maketitle
\pagestyle{plain}
\begin{abstract}
Understanding application resilience (or error tolerance) in the presence of hardware transient faults on data objects is critical to ensure computing integrity and enable efficient application-level fault tolerance mechanisms. However, we lack a method and a tool to quantify application resilience to transient faults on data objects. The traditional method, 
random fault injection, cannot help, because of losing data semantics and insufficient information on how and where errors are tolerated.
In this paper, we introduce a method and a tool (called ``MOARD'') to model and quantify application resilience to transient faults on data objects. Our method is based on systematically quantifying 
error masking events caused by application-inherent semantics and program constructs. We use MOARD to study how and why 
errors in data objects can be tolerated by the application. We demonstrate tangible benefits of using MOARD to direct a fault tolerance mechanism to protect data objects.
\end{abstract}

\IEEEpeerreviewmaketitle

\section{Introduction}
\label{sec:intro}
Transient faults due to high energy particle strikes, wear-out, etc. are expected to become a critical contributor to in-field system failures of high performance computing (HPC). 
If those faults manifest in architecturally visible states (e.g., registers and the memory) and those states hold values of a data object,
then we have transient faults 
on the data object. 
Transient faults 
on a data object impact application outcome correctness. Understanding application resilience to transient faults on data objects is critical to ensure computing integrity in future large scale systems.

Furthermore, many common application-level fault tolerance mechanisms
focus on data objects. Understanding application resilience to transient faults on data objects can be helpful to direct those mechanisms.
Application-level checkpoint is an example of such an application level fault tolerance mechanism. By periodically saving correct
values of some data objects into persistent storage, application-level checkpoint makes application resumable when a failure happens.
Some algorithm-based fault tolerance methods~\cite{ftcg_ppopp13, ft_lu_hpdc13} are other examples. They can detect and locate errors in specific
data objects.
However, those application-level fault tolerance mechanisms can be expensive (e.g., 35\% performance overhead in~\cite{ftfactor_ppopp12}).
If data corruptions of a data object
are easily tolerable by the application, then we do not need to apply those mechanisms to protect the data object, which will improve performance and energy efficiency. 
Hence, understanding application resilience to transient faults on data objects is useful to direct those application level fault tolerance mechanisms. 

However, we do not have a method or a tool 
to quantify application resilience to transient faults on data objects.
The current common practice to understand application resilience
to transient faults in HPC is application-level random fault injection (RFI)~\cite{europar14:calhoun, bifit:sc12, sc16_gpu_fault_injection}. Although RFI is useful, it cannot study application resilience to transient faults on data objects because of the following two reasons.

First, RFI loses application semantics (data semantics). 
RFI randomly selects instructions and triggers random bit flip in 
input or output operands of the instructions.
Typically RFI performs a large amount of random fault injection tests, and then calculates that among all fault injection tests, how many of them succeed (i.e., having correct application outcomes). 
However, we do not know the data value corrupted by RFI belongs to which data object. 
Second, RFI gives us little knowledge on \textit{how} and \textit{where} errors are tolerated~\cite{asplos12:hari}. 
Understanding ``how'' and ``where'' is necessary to identify 
why the application is vulnerable to the value corruption of some data objects, and provides feedback on how to apply application-level fault tolerance mechanisms effectively and efficiently.

In this paper, we introduce a method to model and quantify application resilience to transient faults on data objects. Our method is based on an observation that, 
application resilience to transient faults on data objects is mainly because of application-inherent 
semantics and program constructs. 
For example, a corrupted bit in a data structure could be overwritten by an assignment operation, hence does not cause an outcome corruption; a corrupted bit of a molecular representation in a Monte Carlo method-based simulation 
may not matter to the application outcome because of the statistical nature of the simulation.
Based on the above observation, the quantification of application resilience to transient faults on data objects is equivalent to quantifying error masking events caused by application-inherent semantics and program constructs, and
associating those events with data objects. 
By analyzing application execution information (e.g., the architecture-independent, LLVM~\cite{llvm_lrm} IR trace), we can accurately capture
those error masking events, and provide insightful analysis on 
how and where an 
error tolerance happens.
Furthermore, analyzing application execution information, we can use memory addresses of data objects and track register allocation to associate data values in registers and memory with data objects.
Such a method introduces data semantics into the analysis.
Quantifying application resilience to transient faults on data objects must address a couple of research problems. First, we have little knowledge of the characteristics of 
error masking events. 
This creates a major obstacle to recognize those events and achieve analytical 
quantification. Second, we do not have a good metric to make the quantification. Simply counting the number of 
error masking events cannot provide a meaningful quantification,
because the number 
can be accumulated throughout application execution.
The fact that a data object has many error masking events does not necessarily mean that the application is resilient to
the value corruption of the data object 
because those events may be only a small portion of the total operations on data objects.
Third, determining the impact of an error
occurrence on the correctness of application outcome is challenging. 
The error can propagate to many data objects. Tracking all of those errors for analysis is prohibitive. In addition, an 
error may not impact the correctness of application outcome because of algorithm semantics in the application. However, recognizing algorithm semantics requires detailed application domain knowledge, which is prohibitive for common users. 

Based on the method of quantifying error masking events, 
we systematically model and quantify application resilience to transient faults on data objects, and address the above problems.
We first characterize error masking events and classify them into three classes: 
operation-level error masking, error masking when error propagation,
and algorithm-level error masking. We further introduce a metric. 
The metric quantifies \textit{how often} error masking happens. 
Based on the metric, the comparison of application resilience to transient faults between different data objects is more meaningful than based on simply counting error masking events.
Our classification of error masking events and the proposed metric are fundamental, because they lay a foundation not only for modeling application resilience to transient faults on data objects, but also for other research, such as the placement of error detectors~\cite{
prdc05:karthik} and application checkpoint~\cite{sc10:moody}. 

Based on our classification and metric, we introduce a model.
Given a data object, our model examines operations in the dynamic instruction trace. For each operation that consumes elements of the data object, the model makes the following inference: if an element consumed by the operation has an error, will the application outcome remain correct?
The inference procedure of the model includes three practical techniques to recognize the three classes of error masking events: (1) detecting operation-level error masking based on operation semantics, (2) tracking error propagation by limiting propagation length for analysis, and (3) detecting algorithm-level error masking based on \textit{deterministic} fault injection.
For (2), limiting propagation length is a technique 
based on the characterization of error propagation. This technique
does not impact our conclusion on error masking while avoiding expensive analysis;
for (3), the deterministic fault injection treats the application as a black box without requiring detailed application domain knowledge. 

In summary, this paper makes the following contributions:
\begin{itemize}[leftmargin=*]
\item
A systematic method and a metric to analytically model application resilience to transient faults on data objects, which is unprecedented;
\item A comprehensive classification of error masking events, and methods to recognize them;
\item An open-sourced system tool, MOARD~\cite{moard}, to model application resilience to transient faults on data objects. 
\item An evaluation of representative, computational algorithms and two scientific applications to reveal how application-level error masking typically happens on data objects;
\item A case study to demonstrate the benefit of using a model-driven approach to direct error tolerance designs.
\end{itemize}

\section{Background}
\label{sec:bac}
In this section, we introduce the fault model and give an introductory description of application-level error masking. 


\subsection{Fault Model}

We focus on transient faults that change the values of data objects.
Those faults are not corrected by hardware (e.g., ECC), propagate through a high level of the system,  and become observable to the application~\cite{HWDec_asplos08}.

In terms of application resilience in the existence of corrupted data values, we focus on \textit{application outcome correctness}. 
The application outcome is deemed correct as long as it is \textit{acceptable}.
Depending on the notion of the acceptance, the outcome correctness can refer to precise numerical integrity 
(e.g., the outcome of a multiplication operation must be numerically precise) 
or refer to satisfying a minimum fidelity threshold (e.g., the outcome of an iterative solver must meet certain convergence thresholds).


\subsection{Error Masking}
Error masking can happen at the application level and hardware level.
The application-level error masking happens because of application inherent semantics and program constructs.
The hardware-level error masking happens because a fault does not corrupt the precise semantics of hardware~\cite{micro03:mukherjee}.

The key of our error tolerance modeling is the application-level error masking. 
We particularly study  \textit{error masking that happens to individual data objects}. 
We consider that when an error happens in a data object (\textit{other data objects remain correct before the error happens}) how the error impacts the application outcome correctness. A data object can be an array or other data structures with many data elements. 
Other than data objects, we do not consider the corruption of other application components (e.g., computing logic).
\textit{Hence, we do not aim to model the error tolerance of all application components but focus on data objects}. 
In addition, we focus on errors happening in data objects and directly consumed by the application. 
Latent errors in data objects (i.e., the errors not consumed by the application) are not considered because they do not matter to the application outcome correctness. 

\section{Error Tolerance Modeling}
\label{sec:modeling}
We start with a classification of application-level error masking
and then introduce a modeling
metric.


\subsection{General Description}
\label{sec:general_bg}
Error masking that happens to data objects has various representations.  
Listing~\ref{fig:general_desc} gives a synthetic example to illustrate those representations.
In this example, we focus on a data object, $par\_A$, 
which is an array. 
We study \textit{error masking that happens to this data object}. 
We examine every statement in the example code.
For each statement, we examine if any element of the data object is involved.
If yes, we examine if there is a data corruption in the element, how the data corruption impacts the result correctness of the statement, and how the data corruption propagates to the successor statements which in turn impact the application outcome correctness.

$par\_A$ is involved in 4 statements (Lines 7, 8, 10 and 13). 
The statement at Line 7 has an error masking event:
if an error happens at $par\_A$, (in particular, the data element $par\_A[0]$, which is consumed by the statement), 
the error can be overwritten by an assignment operation, no matter which bit is flipped in $par\_A[0]$.
The statement at Line 8 has no explicit error masking happen. If an error at $par\_A[2]$ occurs, the error propagates to $c$ by multiplication and assignment operations.
If the error propagates to Line 10 (bit shifting), depending on which bit is corrupted at Line 8 and how many bits are shifted at Line 10,
the corrupted bit can be thrown away or remain.
If the corrupted bit is thrown away, then
the error in $par\_A[2]$ propagating from Line 8 to Line 10 is indirectly masked at Line 10 (not directly masked at Line 8).

Line 13 is an invocation of an algebraic multi-grid solver (AMG) taking $par\_A$ as input. AMG 
treats $par\_A$ as a multi-dimensional grid and can tolerate certain data corruptions in the grid,  because of the algorithm semantics of AMG (particularly, 
AMG's iterative structure that mitigates error magnitude and tolerates incorrectness of numerical results~\cite{mg_ics12}).

\begin{figure}
\begin{lstlisting}[
xleftmargin=.02\textwidth, xrightmargin=0.01\textwidth]
void func(double *par_A, double *par_b, 
          double *par_x) 
{
    double c = 0;
    
    //Pre-processing par_A
    par_A[0] = sqrt(initInfo);
    c = par_A[2]*2;
    if (c>THR) 
      par_A[4] = (int)c >> bits; //bit shifting
    
   //Using the algebraic multi-grid solver
    AMG_Solver(par_A, par_b, par_x);
}
\end{lstlisting}
\vspace{-20pt}
\caption{The example code to show error masking that happens to a data object, $par\_A$.} \label{fig:general_desc}
\vspace{-20pt}
\end{figure}

This example reveals many interesting facts.
In essence, a program can be regarded as a combination of data objects and
operations performed on the data objects.
An operation (defined at LLVM instruction level) refers to arithmetic computation, assignment, logical and comparison instructions  
or an invocation of an algorithm implementation. 
An operation may inherently come with error masking effects, exemplified at Line 7 (error overwriting);
an operation may propagate errors, exemplified at Line 8. 
Different operations have different error masking effects, and hence
impact the application outcome differently.
Based on the above discussion, we classify application-level   error masking 
into three classes.

(1) \textbf{Operation-level error masking.} 
An error that happens to the target data object is masked because of the semantics of the operation. Line 7 in
Listing~\ref{fig:general_desc} is an example.

(2) \textbf{Error masking when error propagation.} 
Some error masking events are implicit and have to be identified beyond a single operation. 
In particular, a corrupted bit in a data object is not masked in the current operation (e.g., Line 8 in 
Listing~\ref{fig:general_desc})
but the error propagates to another data object (e.g., the variable $c$) and masked in another operation (e.g., Line 10).
Note that simply relying on isolated operation-level analysis without the error propagation analysis is not sufficient to recognize these error masking events.

(3) \textbf{Algorithm-level error masking.}
Identification of some error masking events must include algorithm-level information.
The identification of these events is beyond the first two classes.
Examples of such events include 
the multigrid solver~\cite{mg_ics12}
and certain sorting algorithm~\cite{prdc13:sharma}.  
The algorithm-level error masking can tolerate errors
that happen to many variables. For example, 
the multigrid solver can tolerate certain errors in hundreds of variables~\cite{mg_ics12}. 
The essence of algorithm-level error masking is typically due to algorithm specific definition on execution fidelity 
and specific program constructs %
that mitigate error magnitude during application execution~\cite{onward10:rinard}.
Limited analysis at individual operations or error propagation
is not sufficient to build up a big picture to capture the algorithm-level fault tolerance.  

Our modeling is analytical and 
relies on the quantification of the above error masking events on data objects.
We create a metric to quantify those events.


\subsection{aDVF: A New Metric}
\label{sec:metric}
To quantify application resilience to transient faults on a data object, the key is to quantify how often error masking happens to the data object. 
We introduce a new metric, \textit{aDVF} (i.e., the application-level Data Vulnerability Factor), to quantify 
application resilience to transient faults on data objects.
aDVF is defined as follows. 

For an operation with the participation of the target data object (maybe multiple data elements of the target data object), we reason that if an error happens to a participating data element of the target data object, 
the application outcome could or could not remain correct in terms of the outcome value and algorithm semantics. 
If the error does not cause an incorrect application outcome,
then an error masking event happens to the target data object.
A single operation can operate on multiple data elements of the target data object. 
For example, an ADD operation can use two elements of the target data object as operands.
For a specific operation, aDVF of the target data object is defined as the total number of error masking events divided by the number of data elements of the target data object involved in the operation.

For example, an assignment operation $a[1] = w$ 
happens to a data object, the array $a$.
This operation involves one data element ($a[1]$) of the target data object $a$.
We calculate aDVF for $a$ in this operation as follows.
If an error happens to $a[1]$, we reason that 
the erroneous $a[1]$ does not impact correctness of the application outcome and the error in $a[1]$ is always masked (no matter which bit of $a[1]$ is flipped). Hence, the number of error masking events for
the target data object $a$ in this operation is 1. Also, the total number of data elements involved in the operation is 1.
Hence, the aDVF value for the target data object in this assignment operation is $1/1=1$.

Based on the above discussion, the definition of aDVF for a data object $X$ in an operation ($aDVF^{X}_{op}$)
is formulated in Equation~\ref{eq:dvf}, where 
$x_i$ is a data element of the target data object $X$ involved in the operation and $m$ is the number of data elements involved in the operation;
$f$ is a function to count error masking events that can happen to a data element. 

\begin{equation} 
\label{eq:dvf}
\footnotesize
	aDVF^{X}_{op} = \sum_{i=0}^{m-1}f(x_i)/m
\end{equation}


To calculate aDVF for a data object in a code segment, we examine operations in
the code segment one by one; 
For each operation that involves any element of the target data object, 
we consider that if a transient fault happens to the element, how many error masking events can happen.
In general, the definition of aDVF for a data object in a code segment is similar to the above for an operation, except that
$m$ is the number of data elements of $X$
involved in all operations of the code segment. \footnote{If a data element is referenced multiple times in the code segment, this data element is counted multiple times in $m$.}
According to the above definition, a higher aDVF value for a data object indicates that the application is more resilient to transient faults on the data object;
Also, an aDVF value should be in $[0,1]$.

To further explain it, we use 
a code segment from LU benchmark in SNU\_NPB benchmark suite 1.0.3 (a C-based implementation of the Fortran-based NAS benchmark suite~\cite{nas}), shown in 
Listing~\ref{fig:advf_example}.

\textbf{An example from LU.} We calculate aDVF for the array $sum[]$. 
Statement $A$ has an assignment operation involving one data element ($sum[m]$) and one   error masking event (i.e., if an error happens to $sum[m]$, 
the error is overwritten by the assignment). Considering that there are five iterations in the first loop ($iter_{num1} = 5$), there are five error masking events  happening to five data elements of $sum[]$.

Statement $B$ has two operations related to $sum[]$ (i.e., an assignment and an addition). The assignment operation involves one data element ($sum[m]$) and has no error masking because the new value is added to $sum[m]$ (not overwriting it); The addition operation involves one data element ($sum[m]$) and may have one error masking (i.e., certain corruptions in $sum[m]$ can be ignored, if ($v[k][j][i][m]*v[k][j][i][m]$) is significantly larger than $sum[m]$). This error masking is counted as $r^\prime$ ($0 \leq r^\prime \leq 1$), depending on the corrupted bit position in $sum[m]$ and the error propagation result (see Sections~\ref{sec:statement_analysis} and~\ref{sec:impl} for further discussion).
In the loop structure where Statement $B$ is, there are ($r^\prime * iter_{num2}$) error masking events that happen to ($2 * iter_{num2}$) elements of $sum[]$, where ``$r^\prime$'' comes from the addition operation~\footnote{The addition operation with the corrupted $sum[m]$ can propagate the error to the assignment. This error propagation effect is included in $r^\prime$.}, 
and $iter_{num2}$ is the number of iterations in the second loop.

\begin{figure}
\begin{lstlisting}[xleftmargin=.02\textwidth, xrightmargin=.01\textwidth]
void l2norm(int ldx, int ldy, int ldz, int nx0, \
 int ny0, int nz0, int ist, int iend, int jst, \
 int jend, double v[][ldy/2*2+1][ldx/2*2+1][5], \
    double sum[5])
{
	int i, j, k, m;
    for(m=0;m<5;m++) //The first loop
      sum[m]=0.0;  //Statement A
    
    for(k=1;k<nz0-1;k++)  //The second loop
      for (j=jst;j<jend;j++)
        for (i=ist;i<iend;i++)
          for (m=0;m<5,m++)
            sum[m]=sum[m]+v[k][j][i][m] \
                  *v[k][j][i][m]; //Statement B

    for(m=0;m<5;m++){  //The third loop
      sum[m]=sqrt(sum[m]/((nx0-2)*  \
        	(ny0-2)*(nz0-2))); //Statement C
    }
} 
\end{lstlisting}
\vspace{-20pt}
\caption{A code segment from LU.} 
\vspace{-20pt}
\label{fig:advf_example} 
\end{figure}

Statement $C$ has two operations 
related to $sum[]$ (i.e., an assignment and a division) but only the assignment operation has error masking (overwriting).
In the loop structure where Statement $C$ is, there are five iterations ($iter_{num3} = 5$).
Hence, there are five error masking events that happen on five data elements of the target data object. In 
summary, the aDVF calculation for $sum[]$ is 

\begin{equation}
\label{eq:dvf_exp}
\footnotesize
aDVF_{op}^{sum}= 
\frac{1*iter_{num1}+r^{\prime}*iter_{num2}+1*iter_{num3}}{1*iter_{num1}+(1+1)*iter_{num2}+(1+1)*iter_{num3}},
\end{equation}

where each term in the numerator is the number of error masking events in the first, second, and third loop, respectively; 
each term in the denominator
is the number of target data elements involved in each loop; $iter_{num1}=5, iter_{num3}=5$ and $iter_{num2} = (nz0-2)*(jend-jst)*(iend-ist)*5$.

To calculate aDVF for a data object, we must rely on effective identification and counting of error masking events (i.e., the function $f$).
In Sections~\ref{sec:statement_analysis},~\ref{sec:fault_propagation_analysis} and ~\ref{sec:algo_analysis}, 
we introduce a series of counting methods based on the classification of error masking events. 

\subsection{Operation-Level Analysis}
\label{sec:statement_analysis}
To identify error masking events at the operation level, we analyze 
all possible operations. 
In particular, we analyze 
architecture-independent, LLVM instructions 
and characterize their error tolerance based on operation semantics.
We classify the operation-level error masking as follows. 

(1) \textbf{Value overwriting}.  
An operation writes a new value into a data element of the target data object and the 
error in the data element (no matter where the corrupted bit is in the data element) is masked. 
For example, the store operation overwrites the error in the store destination. 
We also include \textit{trunc} and bit-shifting operations into this category because the error could be truncated or shifted away in those operations. 

(2) \textbf{Logical and comparison operations}.
If an error in the target data object does not
change the correctness of logical and comparison operations, the error is masked.  
Examples of such operations 
include logical \textit{AND} and the predicate expression in a \textit{switch} statement.

(3) \textbf{Value overshadowing}.
If the corrupted data value in an operand of an addition or subtraction operation 
is overshadowed by the other correct operand involved in the operation,
then the corrupted data can have an 
ignorable impact on the correctness of application outcome. 
For example, the data value ``10'' in an addition operation (``10e+6 + 10'') is corrupted and the addition operation becomes ``10e+6 + 11''. But such data corruption may not matter to the application outcome because the operand ``10e+6'' is much larger than the magnitude of the data corruption.
We further discuss how the overshadowing effect is determined in Section~\ref{sec:impl}.

The above three operation-level error masking impacts the application outcome differently. 
Error masking based on value overwriting and logic and comparison operations can make the application outcome numerically the same as the error-free case. Error masking based on value overshadowing can make the application outcome numerically different from or the same as the error-free case.

For value overshadowing, if the application outcome is numerically different, the application outcome can still be acceptable because of algorithm semantics; if the application outcome is numerically the same, operations \textit{after} the value overshadowing must 
help tolerate corrupted bits. 
For the above two cases, we do not attribute error masking to the algorithm level or error propagation level. Instead, we attribute it to operation-level value overshadowing because value overshadowing initiates error masking. Without value overshadowing, algorithm or error propagation may not mask errors. 

The effectiveness of the above error masking heavily relies on the error pattern. \textit{The \textbf{error pattern} is defined by how erroneous bits are distributed within a corrupted data element} (e.g., single-bit vs. spatial multiple-bit, least significant bit vs. most significant bit).  
Depending on where the erroneous bit is, the error in the data object could or could not be masked. 
Take as an example the bit shift operation (Line 10) in Listing~\ref{fig:general_desc}.
Depending on the error pattern, 
the shift operation can remove or keep the corrupted bit.

To determine the existence of the above (2) and (3) error masking,
we must consider error patterns (i.e., the spatial aspect of errors~\cite{ipdps16:vishnu}).
In the practice of our resilience modeling, 
given an operation to analyze,
we enumerate possible error patterns for the target data object. 
Then, we derive the existence of error masking for each error pattern without application execution. Suppose there are $n$ error patterns and $m$ ($0 \leq m \leq n$) of which have error masking.
Then the number of error masking events is calculated as $m$/$n$,
which is a statistical quantification of possible error masking.
In the example of the bit shift (Line 10 in Listing~\ref{fig:general_desc}), assuming that $c$ is 64-bit and we consider single-bit errors, then there are 64 error patterns. For each error pattern, we 
decide if the corrupted bit is shifted away. If 10 of the 64 fault patterns have the corrupted bit shifted, then the number of error masking events for the data object $c$ in this shift operation is 10/64. 


\subsection{Error Propagation Analysis}
\label{sec:fault_propagation_analysis}
If we analyze a specific error pattern in an operation (named ``target error pattern'' and ``target operation'' in the rest of this section) and determine that the error cannot be masked in the target operation, 
then we use error propagation analysis to capture error
masking (i.e., the temporal aspect of errors~\cite{ipdps16:vishnu}).
Using a dynamic instruction trace as input, the error propagation analysis tracks whether the errors (including the original one and the new ones because of error propagation) are masked in the successor operations based on the operation-level analysis without application execution. If all of the errors are masked and hence 
the application outcome remains \textit{numerically the same} as the error-free case, 
then we claim that the original error in the target operation is masked. 

For the error propagation analysis, a big challenge is to 
track all contaminated data which can quickly increase as the error propagates. 
Tracking all the contaminated data significantly increases analysis time and memory usage. 
A solution to this challenge is \textit{deterministic} fault injection. Different from random fault injection, the deterministic
fault injection injects an error at the target operation using the target error pattern and then run the application to completion. 
If the application outcome is \textit{numerically the same} as the error-free case, then the original and the new errors are masked, and the error masking based on error propagation takes effect. 
If the application outcome is numerically different but still accepted, then the algorithm-level error masking takes effect.  

Because of the deterministic fault injection, we do not need to analyze operations one by one to track data flow and error contamination. Hence it is faster. However, the deterministic fault injection can still be time-consuming, if application execution time is long. To improve the efficiency of the error propagation analysis, we optimize the analysis based on the characteristics of error propagation. 

\textbf{Optimization: bounding propagation path.} 
We observe that tracking a limited number of operations ($k$ operations) after the target operation is often sufficient to decide the existence of the propagation-based error masking. Our observation is based on 1000 random fault injection tests on 16 data objects from eight benchmarks (see Table~\ref{tab:benchmark} for benchmark details). We observe that 87\% of the fault injection tests that cannot mask errors within 10 operations ($k=10$) after fault injection lead to numerically incorrect application outcomes;
100\% of the fault injection tests that cannot mask errors within 50 operations ($k=50$) after fault injection lead to numerically incorrect application outcomes. This fact indicates that errors that are not masked within a limited number of operations have little chance to be masked by further error propagation. 

The rationale to support the above observation is as follows. 
An error in a data object typically propagates
to a large amount of data (objects) quickly. 
After a certain number of operations, it is very unlikely that all errors are able to be masked by further error propagation and making a conclusion of no error masking \textit{by error propagation} is correct in most cases. 

Based on the above observation, we only need to track the first $k$ operations after the target operation to determine the existence of the propagation-based error masking. In particular, after analyzing $k$ operations ($k=50$ in our evaluation), 
(1) If not all errors due to error propagation are masked at the operation level, we conclude that the errors will not be masked at the operation level by further error propagation.
But those errors may be masked by algorithm (if the user wants to do algorithm-level analysis), pending further investigation;
(2) If all errors due to error propagation are masked and based on the operation-level analysis we can derive that the application outcome remains numerically correct, 
then we claim error masking due to error propagation happens.

\subsection{Algorithm-Level Analysis}
\label{sec:algo_analysis}
Identifying the algorithm-level error masking demands domain and algorithm knowledge.  In our modeling, we want to minimize the usage of that knowledge, 
such that the modeling methodology can be general across different domains.
The traditional random fault injection treats the program as a black-box.  Hence, using the traditional random fault injection could be an effective tool to identify the algorithm-level error masking.
However, to avoid the randomness, we use the deterministic fault injection again.

In particular, when we analyze a specific error pattern in a target operation and decide that the error cannot be masked in the target operation and next $k$ operations, we inject an error using the error pattern in the target operation and run the application to completion. 
If the application outcome is numerically different from the error-free case but acceptable in terms of algorithm semantics, then algorithm-level error masking takes effect.
If the application outcome is numerically the same, then error masking due to error propagation happens, which should be rare based on the above discussion on ``bounding propagation path''.

\textbf{Discussion}: Although we employ the deterministic fault injection, it cannot replace our modeling because of two reasons. First, the fault injection space without our modeling is typically huge (trillions of fault injection sites~\cite{asplos12:hari}), which is prohibitive for implementation. Second, the deterministic fault injection tells us little about how an error is tolerated.  

\section{Implementation}
\label{sec:impl}

\begin{figure*}
  \begin{center}
  \includegraphics[height=0.13\textheight,keepaspectratio]{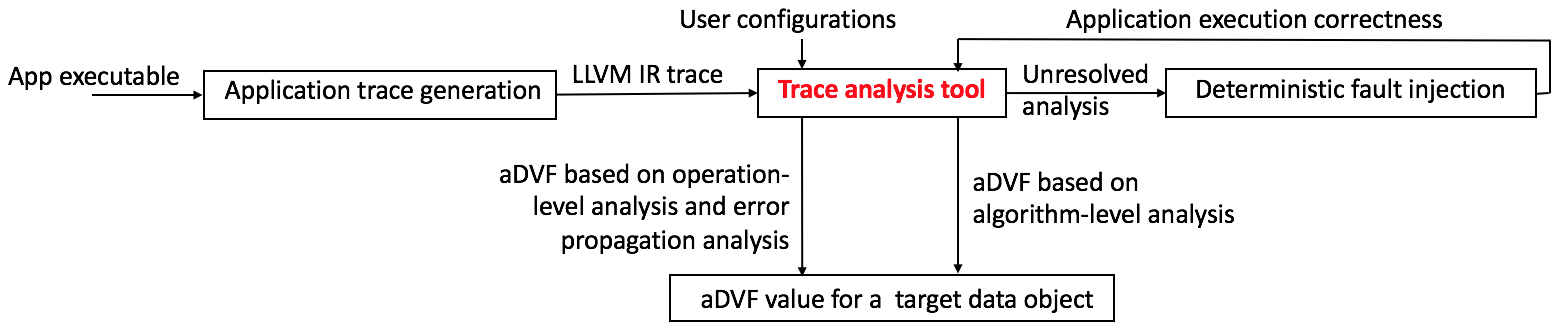}
  \caption{MOARD, a tool for modeling application resilience to transient faults on data objects}
  \label{fig:tool_framework}
  \end{center}
  \vspace{-20pt}
\end{figure*}

To calculate the aDVF value for a data object, 
we develop a tool, named~\textit{MOARD} (standing for \textit{MO}deling \textit{A}pplication \textit{R}esilience to transient faults on  \textit{D}ata data objects). 
Figure~\ref{fig:tool_framework} shows the tool framework and its algorithm.
MOARD has three components: an application trace generator, a trace analysis tool, and a deterministic fault injector.

The \textbf{application trace generator} is an LLVM instrumentation pass to generate a dynamic LLVM IR trace.
LLVM IR is architecture independent and each instruction in the dynamic IR trace corresponds to one operation.
We extend a trace generator~\cite{ispass13:shao} to enable trace generation for MPI applications. During the trace analysis, we consider error propagation by MPI communication, but do not consider those cases where errors happen in the communication. 

The \textbf{trace analysis tool} is the core of MOARD. Using an application trace as input, the tool can calculate the aDVF value of 
any data object with known memory address range. 
In particular, the trace analysis tool conducts the operation-level and error propagation analysis. 
For those unresolved analyses,
the trace analysis tool will output 
a set of fault injection information for the deterministic fault injection. Such information includes dynamic instruction IDs, IDs of the operands that reference the values of the target data object, 
and the bit locations of the operands that correspond to those error patterns with undetermined error masking.
After the fault injection results 
(i.e., the numerical values of application outcome and whether the outcome is acceptable)
are available from the deterministic fault injector,
we re-run the trace analysis tool, and use the fault injection results to address the unresolved analyses 
and update the aDVF calculation. 

For the error propagation analysis, we associate data semantics (the data object name) with the data values in registers,
such that we can identify the data of the target data object in registers. 
To associate data semantics with the data in registers, 
MOARD tracks the register allocation when analyzing the trace, such that we can know at any moment which registers have the data of the target data object. 

To \textbf{determine the existence of value overshadowing} in an addition or subtraction operation, 
we use the deterministic fault injection. Particularly, given a
target operand in an addition or subtraction operation for value overshadowing analysis, we enumerate all error patterns for deterministic fault injection tests. 
If the following two conditions are true, then we derive that the value overshadowing happens in the operation: 
\begin{itemize}[leftmargin=*]
\item Some error patterns result in small magnitudes of the operand (smaller than the magnitude of the other operand in the operation); the application outcome is acceptable. 
\item The other error patterns result in larger magnitudes of the operand (larger than those in the first condition) but the application outcome is not acceptable. 
\end{itemize}
The error masking of the value shadowing is quantified as $x/y$, where $x$ is the number of error patterns in the first condition and $y$ is the number of all error patterns.
For example, suppose we have an addition operation ($a+b$, $a=1000$ and $b=1$) and $b$ is our target data object. We enumerate error patterns in $b$ (assuming 32 single-bit-flip error patterns). If five patterns result in the values of $b$ as 0, 3, 5, 9 and 17, which are smaller than $a$ and the application outcome is acceptable, and the other 26 patterns result in larger $b$ (larger than 0, 3, 5, 9, and 17) but the application outcome is not acceptable, then the value overshadowing happens (the corrupted $b$ is overshadowed by $a$), and is quantified as 5/32.

The \textbf{deterministic fault injector} is a tool to 
resolve those error masking analyses undetermined by the trace analysis tool. 
The input to the deterministic fault injector is a list of fault injection sites generated by the trace analysis tool.
Similar to the application trace generation, the deterministic fault injector is also based on the LLVM instrumentation. 
We use the LLVM instrumentation to count dynamic instructions and trigger bit flips. 
The application execution will trigger bit flip when a fault injection site is encountered.

To \textbf{accelerate the calculation of aDVF}, we leverage the existing
work~\cite{asplos12:hari, isca14:hari} that explores ``error equivalence'' based on the similarity of intermediate execution states to avoid repeated analysis and fault injections on instructions. 
During our evaluation, MOARD calculates aDVF for 16 data objects
in eight benchmarks 
within one day on a cluster of 256 cores, which is comparable to the execution time of existing fault injection work~\cite{asplos12:hari, isca14:hari}.

\section{Evaluation}
\label{sec:evaluation}

In this section, we use aDVF as an metric to evaluate application resilience to transient faults on data objects with a set of benchmarks. Furthermore, we validate the accuracy of our aDVF calculation. We also compare aDVF calculation with the traditional fault injection to show the power and benefits of aDVF calculation.

\begin{table*}[!t]
\begin{center}
\caption {Benchmarks and applications for the study}
\label{tab:benchmark}
\scriptsize
\begin{tabular}{|p{1.7cm}|p{7.5cm}|p{4cm}|p{2.5cm}|}
\hline
\textbf{Name} 	& \textbf{Benchmark description} 		& \textbf{Code segment for evaluation}  			& \textbf{Target data objects}             \\ \hline \hline
CG (NPB)             & Conjugate Gradient, irregular memory access (input class S)   & The routine conj\_grad in the main loop  & The arrays $r$ and $colidx$     \\\hline
MG (NPB)    	       & Multi-Grid on a sequence of meshes (input class S)             & The routine mg3P in the main loop & The arrays $u$ and $r$ 	\\ \hline
FT (NPB)             & Discrete 3D fast Fourier Transform (input class S)            & The routine fftXYZ in the main loop  & The arrays $plane$ and $exp1$    \\ \hline
BT (NPB)             & Block Tri-diagonal solver (input class S)         		& The routine x\_solve in the main loop & The arrays $grid\_points$, $u$	\\ \hline
SP (NPB)             & Scalar Penta-diagonal solver (input class S)         		& The routine x\_solve in the main loop & The arrays $rhoi$ and $grid\_points$  \\ \hline
LU (NPB)            & Lower-Upper Gauss-Seidel solver (input class S)        	& The routine ssor 	& The arrays $u$ and $rsd$ \\ \hline \hline
LULESH~\cite{IPDPS13:LULESH} & Unstructured Lagrangian explicit shock hydrodynamics (input 5x5x5) & 
The routine CalcMonotonicQRegionForElems 
& The arrays $m\_elemBC$ and $m\_delv\_zeta$ \\ \hline
AMG2013~\cite{anm02:amg} & An algebraic multigrid solver for linear systems arising from problems on unstructured grids (we use  GMRES(10) with AMG preconditioner). We use a compact version from LLNL with input matrix $aniso$. & The routine hypre\_GMRESSolve & The arrays $ipiv$ and $A$   \\ \hline
\end{tabular}
\end{center}
\vspace{-10pt}
\end{table*}

\subsection{Evaluating Application Resilience to Transient Faults on Data Objects Using aDVF}

We study 12 data objects from six benchmarks of the NAS parallel benchmark (NPB) suite and four data objects from two scientific applications. 
Those data objects are chosen to be representative: they have various data access patterns and participate in different execution phases.  
Table~\ref{tab:benchmark} gives 
details on the benchmarks and applications.
The maximum error propagation path for aDVF analysis is 50, 
for which we do not lose analysis accuracy as we discuss in Section~\ref{sec:fault_propagation_analysis}.
Similar to~\cite{asplos12:hari, isca14:hari, micro16:ventatagiri}, we only study single-bit errors because they are the most common 
errors. 



\begin{figure*}
	\centering
        \includegraphics[width=0.95\textwidth]{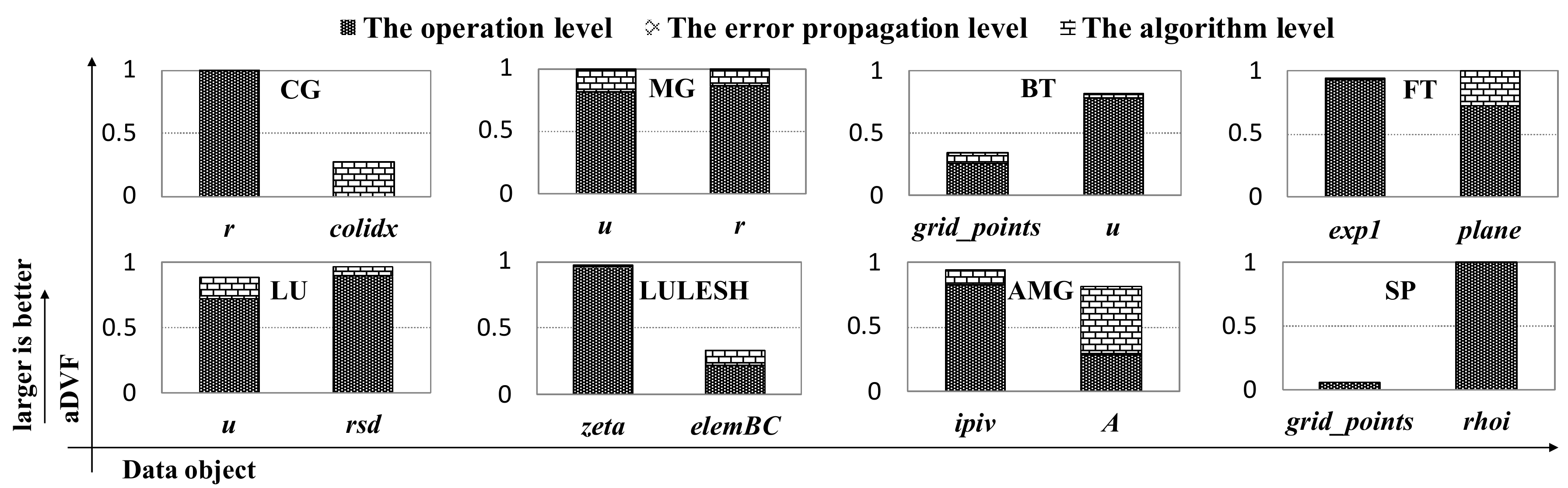}
        \caption{The breakdown of aDVF results based on the three level analysis. The $x$ axis is the data object name.}
        \label{fig:aDVF_3tiers_profiling}
        \vspace{-5pt}
\end{figure*}

Figure~\ref{fig:aDVF_3tiers_profiling}
shows the aDVF results and breaks them down into the three levels 
(i.e., the operation level, error propagation level, and algorithm level).

\begin{figure*}
	\centering
	\includegraphics[width=0.94\textwidth]{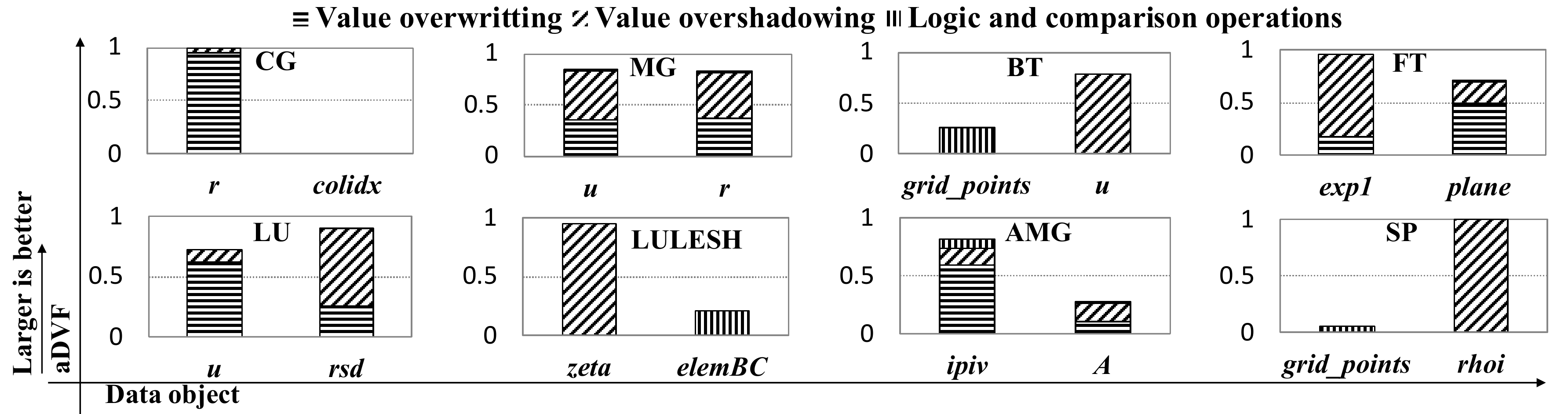}
	\caption{The breakdown of aDVF results based on value overwriting, value overshadowing, and logic and comparison operation at the levels of operation and error propagation. The $x$ axis is the data object name. \textit{zeta} and \textit{elemBC} in LULESH are \textit{m\_delv\_zeta} and \textit{m\_elemBC}.} 
	\label{fig:aDVF_3classes_profiling}
    \vspace{-15pt}
\end{figure*}

Error masking happens commonly in data objects across benchmarks and applications including those scientific applications (e.g., LULESH and AMG) that are highly sensitive to data correctness. 
Several data objects (e.g., $r$ in CG, and $exp1$ and $plane$ in FT)
have aDVF values close to 1 in Figure~\ref{fig:aDVF_3tiers_profiling}, 
which indicates that most operations working on these data objects
have error masking. Those data objects are double-precision floating-point
and their error masking mainly comes from value overshadowing and overwriting (Figure~\ref{fig:aDVF_3classes_profiling}).
However, a couple of data objects have much less intensive error masking.
For example, the aDVF value of $colidx$ in CG is only 0.28 (Figure~\ref{fig:aDVF_3tiers_profiling}). 
Further study reveals that $colidx$ is an integer array to store indexes of sparse matrices and there is few operation-level or error propagation-level error masking  (Figure~\ref{fig:aDVF_3classes_profiling}).
Its corruption can easily cause segmentation error caught by the
deterministic fault injection.
$grid\_points$ in SP and BT also have a small aDVF value (0.06 and 0.38 for SP and BT respectively in Figure~\ref{fig:aDVF_3tiers_profiling}).
Further study reveals that the 
array $grid\_points$ defines input problems for SP and BT.  
An error in $grid\_points$ 
can easily cause major changes in computation
caught by the error propagation analysis. 

\textbf{Evaluation conclusion 1}: The above aDVF-based analysis reveals the variation of application resilience to transient faults on data objects 
and provides insights on whether the corruption on a data object impacts application outcomes, which is useful to direct fault tolerance mechanisms. 

We further notice that the data objects \textit{colidx} and \textit{r} in CG have 2.19e+09 and 4.54e+07 error masking events (not shown in Figure~\ref{fig:aDVF_3tiers_profiling}), respectively. Although \textit{colidx} has more error masking events, CG is not more resilient to errors on \textit{colidx} than on \textit{r}. In particular, 75\% bit flips that happen in the elements of \textit{colidx} involved in the operations of CG causes incorrect application outcome or segmentation faults, while less than 1\% in $r$. The above observation provides a strong support to introduce the metric, aDVF.

\textbf{Evaluation conclusion 2}: Simply counting the number of error masking events is not sufficient to evaluate application resilience to errors on data objects. 

We further look into the results based on the analysis of the three levels. Operation-level error masking is very common.
Figure~\ref{fig:aDVF_3tiers_profiling} shows that there are 12 data objects whose operation-level error masking contributes more than 70\% of the aDVF values. For $exp1$ in FT and $rhoi$ in SP, the contribution of the operation-level error masking is close to 99\%.

We further notice that the contribution of error masking at the error propagation level to the aDVF result is very limited.
For most of the data objects, the contribution 
is less than 10\% (Figure~\ref{fig:aDVF_3tiers_profiling}).
For five data objects ($colidx$ in CG, $grid\_points$ and $u$ in BT, and  $grid\_points$ and $rhoi$ in SP),  there is no such error masking. Note that our analysis at the error propagation level is valid even if we increase the error propagation length. We discuss the impact of error propagation length in Section~\ref{sec:fault_propagation_analysis}.

Different from error masking at the error propagation level, the contribution of the algorithm-level error masking to the aDVF result is relatively large. For example, the algorithm-level error masking contributes 19\% to the aDVF value for $u$ in MG and 27\% for $plane$ in FT (Figure~\ref{fig:aDVF_3tiers_profiling}).
The large contribution for $u$ in MG is consistent with the existing work~\cite{mg_ics12}. 
For FT (particularly 3D FFT), the large contribution of algorithm-level error masking in $plane$ comes from frequent transpose and 1D FFT computations that average out 
the data corruption.
CG, as an iterative solver, is known to have the algorithm-level error masking
because of the iterative nature~\cite{2-shantharam2011characterizing}.
Interestingly, the algorithm-level error masking in CG contributes most to 
application resilience to transient faults on $colidx$ 
which is a vulnerable integer data object (Figure~\ref{fig:aDVF_3tiers_profiling}).

\textbf{Evaluation conclusion 3}: The aDVF analysis gives us deep information on how errors are tolerated.
This may be useful for refactoring application (e.g., using different algorithms or different data structures and data types) to improve error tolerance of data objects. 

We further break down the aDVF results based on classifications of the value overwriting, logical and comparison operations, and value overshadowing) based on the analysis at the operation and error propagation levels, shown in 
Figure~\ref{fig:aDVF_3classes_profiling}. 
We have the following observation.

The value overshadowing is very common, 
especially for (double-precision) floating point data objects (e.g., $u$ in BT, $zeta$ in LULESH, and $rhoi$ in SP in Figure~\ref{fig:aDVF_3classes_profiling}).
This finding has an important indication for studying application-level error tolerance. We have the following conclusion:
the impact of data corruption
can be correlated with the input problem,
because different input problems can have different values of the data objects, which in turn have different effects of value overshadowing.
Hence, the existing conclusions on application-level fault tolerance~\cite{europar14:calhoun, bifit:sc12, sc16_gpu_fault_injection, prdc13:sharma, isca07:li} with single input problems must be re-examined with different input problems to validate the conclusions of application resilience.
\vspace{-5pt}
\subsection{Model Validation}
\label{sec:model_validation}

In this section, we aim to (1) validate the accuracy of our approach to calculate aDVF, and (2) demonstrate that aDVF correctly quantifies application resilience to transient faults on \textit{data objects}.
 
We validate our modeling approach by comparing the aDVF result with the result of \textit{exhaustive fault injection} (particularly, the success rate of exhaustive fault injection tests). The exhaustive fault injection is different from the traditional random fault injection. With an exhaustive fault injection campaign, we inject faults into \textit{all} valid fault injection sites. A valid fault injection site is a bit in an instruction operand or output that has a value of the target data object. We use those fault injection sites, because we quantify application resilience to transient faults on \textit{data objects}. The exhaustive fault injection is accurate to quantify application resilience to transient faults on data objects, because of its full coverage of all fault sites. However, the number of valid fault injection sites can be very large (e.g., trillions of sites in CG (Class A)). Hence, although the exhaustive fault injection is accurate and good for model validation in this section, this method is not practical, compared with aDVF.


Note that the aDVF result \textit{cannot be exactly the same} as the exhaustive fault injection result, because the definitions of aDVF and exhaustive fault injection are different. Hence, we validate the modeling accuracy by quantifying application resilience to transient faults for multiple data objects, and then ranking them based on the quantification. Ideally, the rank order of data objects based on the aDVF calculation should be exactly the same as that based on the exhaustive fault injection. A correct order of data objects in terms of
application resilience to transient faults is critical to decide which data objects should be protected by fault tolerance mechanisms.

\begin{figure}
		\begin{center}
		\includegraphics[width=0.4\textwidth]{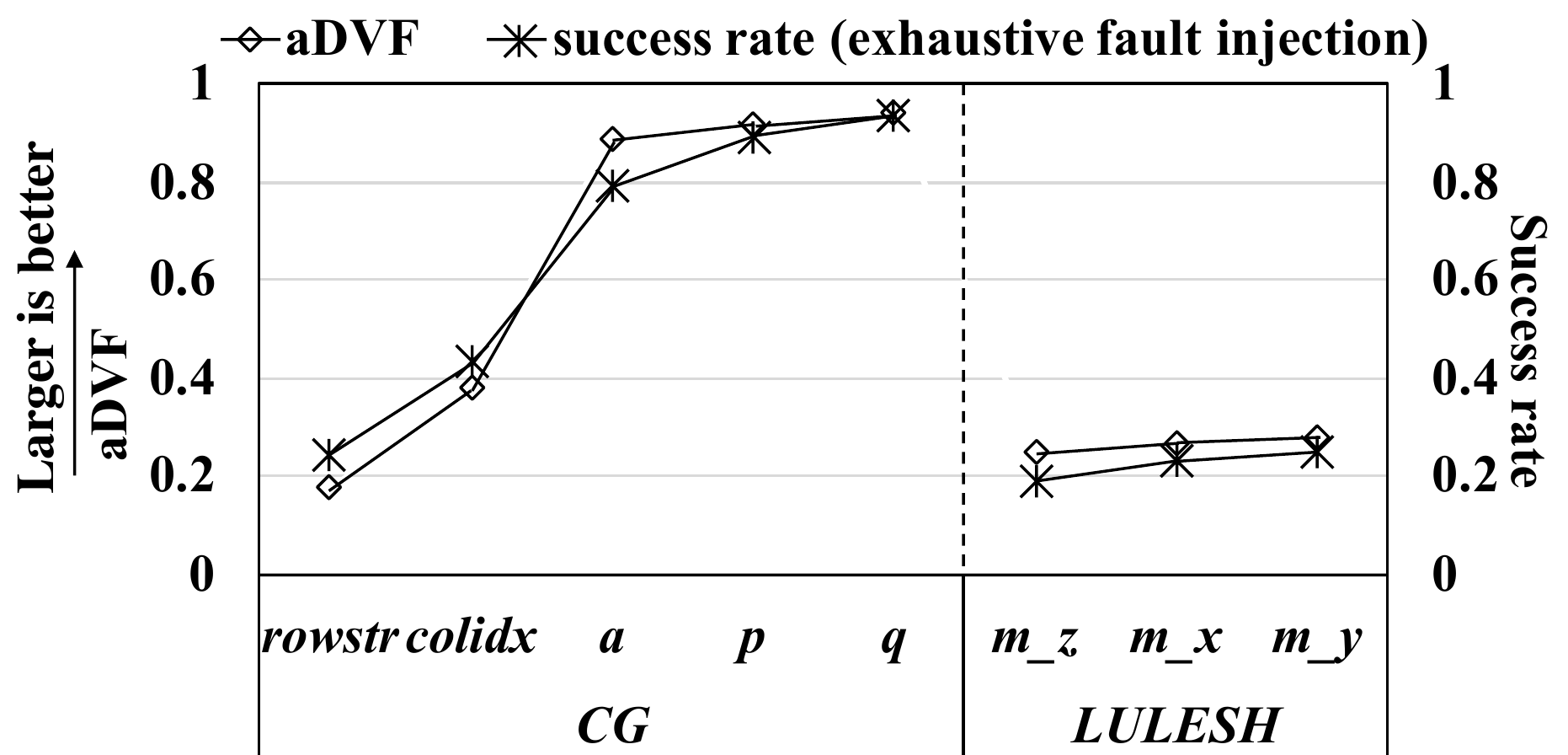}
		\caption{Model validation against exhaustive fault injection. The x axis shows the data object name.}
		\label{fig:model_validation}
		\end{center}
        \vspace{-15pt}
\end{figure}

We focus on a function ($conj\_grad()$) from CG and a function ($CalcMonotomicQRegionForElems()$) from LULESH. We study major data objects in the two functions (those data objects take most of memory footprint). We use single-bit flip in fault injection. The results are shown in Figure~\ref{fig:model_validation}. We notice that the aDVF and exhaustive fault injection rank the data objects in the same order. aDVF correctly reflects application resilience to transient faults on data objects.



\subsection{Comparing aDVF Calculation with the Traditional Random Fault Injection (RFI)} 

We compare aDVF calculation with RFI. We aim to reveal the limitation of this traditional approach, and demonstrate the predictive power of aDVF, compared to RFI.

\subsubsection{\textbf{RFI}}
We use the following method for RFI. We use valid fault injection sites, as defined in Section~\ref{sec:model_validation}, for RFI. In each fault injection test, we randomly trigger a single-bit flip in a valid fault injection site. The number of fault injection tests is determined by a statistical approach~\cite{date09:leveugle} using confidence-level of 95\% to ensure statistical significance. We do seven sets of fault injection tests, and the number of fault injection tests in the seven sets ranges from 500 to 3500 with a stride of 500. We use three equal-sized, floating-point arrays ($m\_x$, $m\_y$, and $m\_z$) in the function $CalcMonotomicQRegionForElems()$ of LULESH for study.

\begin{figure}
	\begin{center}
		\includegraphics[height=0.18\textheight]{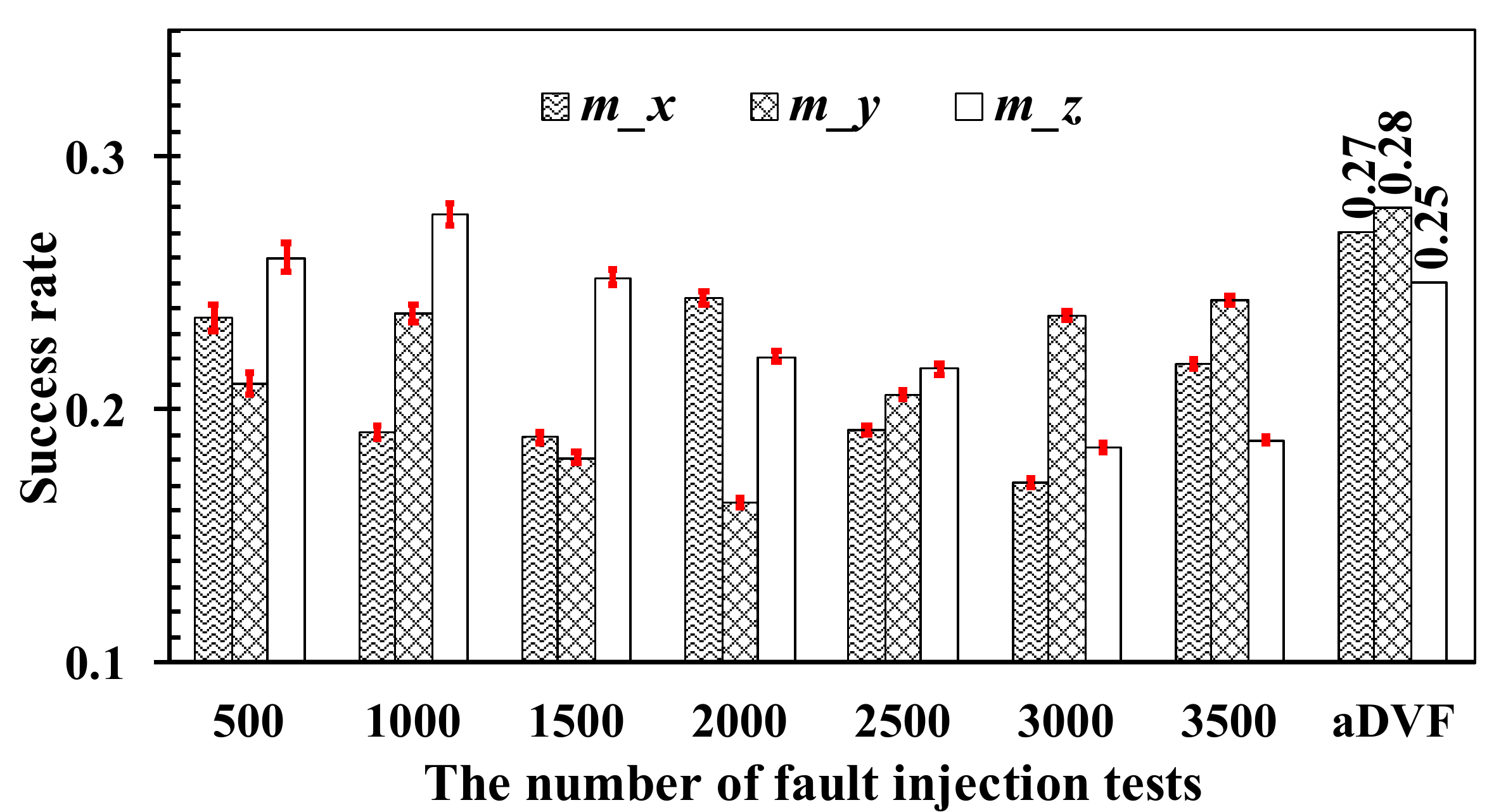} 
		\caption{The RFI results with the margin of error (the confidence level 95\%) and aDVF results. The results are for three data objects ($m\_x$, $m\_y$, and $m\_z$) from $CalcM onotomicQRegionF orElems()$ of LULESH.}
        \vspace{-20pt}
		\label{fig:fi_randomness}
	\end{center}
\end{figure}


Figure~\ref{fig:fi_randomness} shows the results of RFI (the success rate). \textit{The figure also shows the margin of error} (shown as small red bars in the figure).
The results reveal that the results of RFI are sensitive to the number of fault injection tests. For example, for $m\_z$, the success rates of RFI are 0.28 and 0.19 for 1000 and 3000 random fault injection tests, respectively. There is 49\% difference between the two results. Furthermore, in terms of application resilience to transient faults on data objects, we cannot rank the three target data objects in a consistent order across the seven test sets. For example, the success rate of RFI for $m\_x$ is lower than that for $m\_z$, when the number of fault injection tests is 500, 1000, and 1500. However, the observation is opposite, when the number of fault injection tests is 2000 and 3500. In other words, using RFI, we cannot make any conclusion that LULESH is more resilient to transient faults on a data object than on another data object (even through the margin of error is considered). The reason is three-fold:  randomness of RFI, limited confidence level, and inability to capture error masking events.

\subsubsection{\textbf{aDVF}}
We measure aDVF of the three data objects. Figure~\ref{fig:fi_randomness} shows the results (see the last group of bars). We rank the three objects in a determined order (i.e., no inconsistence in the aDVF calculation results, no matter how many times we calculate aDVF). The order is also verified by the accurate, exhaustive fault injection (see Section~\ref{sec:model_validation} for discussion). Having a determined order is important for guiding error tolerant designs (e.g., deciding which data object should be protected by a fault tolerance mechanism).



\textbf{Evaluation conclusion 4:} The calculation of aDVF is deterministic, meaning that we can deterministically rank data objects in terms of application resilience to transient faults on the data objects. Using the traditional RFI, we cannot do so. RFI can be ineffective for guiding error tolerant designs.

\section{Case Study}
\label{sec:case_study}
In this section, we study a case of using aDVF to help system designers decide whether a specific application-level fault tolerance mechanism is helpful to improve application resilience to transient faults on data objects.

Application-level fault tolerance mechanisms, such as algorithm-based fault tolerance~\cite{21-chen2011algorithm,jcs13:wu,tc84_abft}, 
are extensively studied as a means to increase application resilience to transient faults on data objects. However, those mechanisms
can come with big performance and energy overheads 
(e.g., 35\% performance loss in~\cite{ftfactor_ppopp12}). 
To justify the necessity of using those mechanisms, we must quantify the effectiveness of those mechanisms. 
With the introduction of aDVF, we can evaluate if 
application resilience to transient faults on data objects
is effectively improved with fault tolerance mechanisms in place.

\begin{figure}
	\centering
	\includegraphics[width=0.5\textwidth]{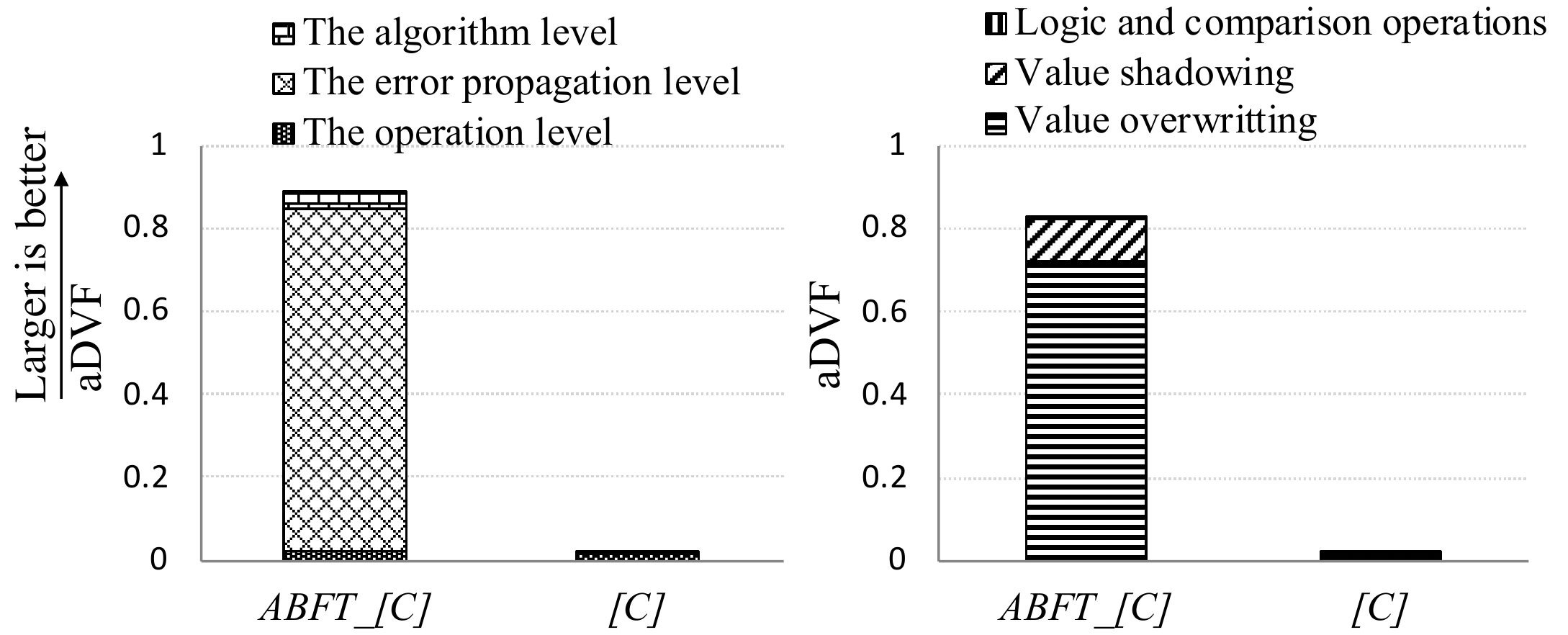}
	\caption{Using aDVF analysis to study application resilience to transient faults on $C$ in matrix multiplication (MM). Notation: [C] is MM without applying ABFT on $C$; ABFT\_[C] is MM with ABFT taking effect.}
	\label{fig:abft_advf}
    \vspace{-15pt}
\end{figure}

We focus on a specific application-level fault tolerance mechanism,
the algorithm-based fault tolerance (ABFT) for general matrix multiplication ($C=A \times B$)~\cite{jcs13:wu}.
This ABFT mechanism encodes matrices $A$, $B$, and $C$ into a new form with checksums. If an error happens in an element of $C$, leveraging the checksums, we are able to correct and detect the erroneous element.  
We apply the aDVF analysis on this ABFT and the matrix $C$ is the target data object. We compare the aDVF values of $C$
with and without ABFT. Figure~\ref{fig:abft_advf} shows the results.
The figure shows that ABFT effectively improves error tolerance of $C$: the aDVF value increases
from 0.0172 to 0.82 (the larger is better). The improvement mostly comes from the value overwriting during error propagation.
This result is expected because a corrupted element of $C$ is not corrected by ABFT right away.
Instead, it will be corrected in a specific verification phase of ABFT during error propagation.

Given the effectiveness of this ABFT, we further explore whether this ABFT can help us improve resilience to transient faults on a data object in an application, Particle Filer (PF) from Rodinia~\cite{Che+:IISWC09}, without knowing the application resilience of PF.
PF has a critical variable, $xe$, which is repeatedly used to store vector multiplication results. Given the fact that a vector can be treated as a special matrix, we can apply ABFT to protect $xe$ for those vector multiplications.
Using $xe$ as our target data object, we perform the aDVF analysis with and without ABFT. We want to answer a question: \textbf{Will using ABFT be an effective fault tolerance mechanism for protecting $xe$ in PF}?

\begin{figure}
	\centering
	\includegraphics[width=0.5\textwidth]
   {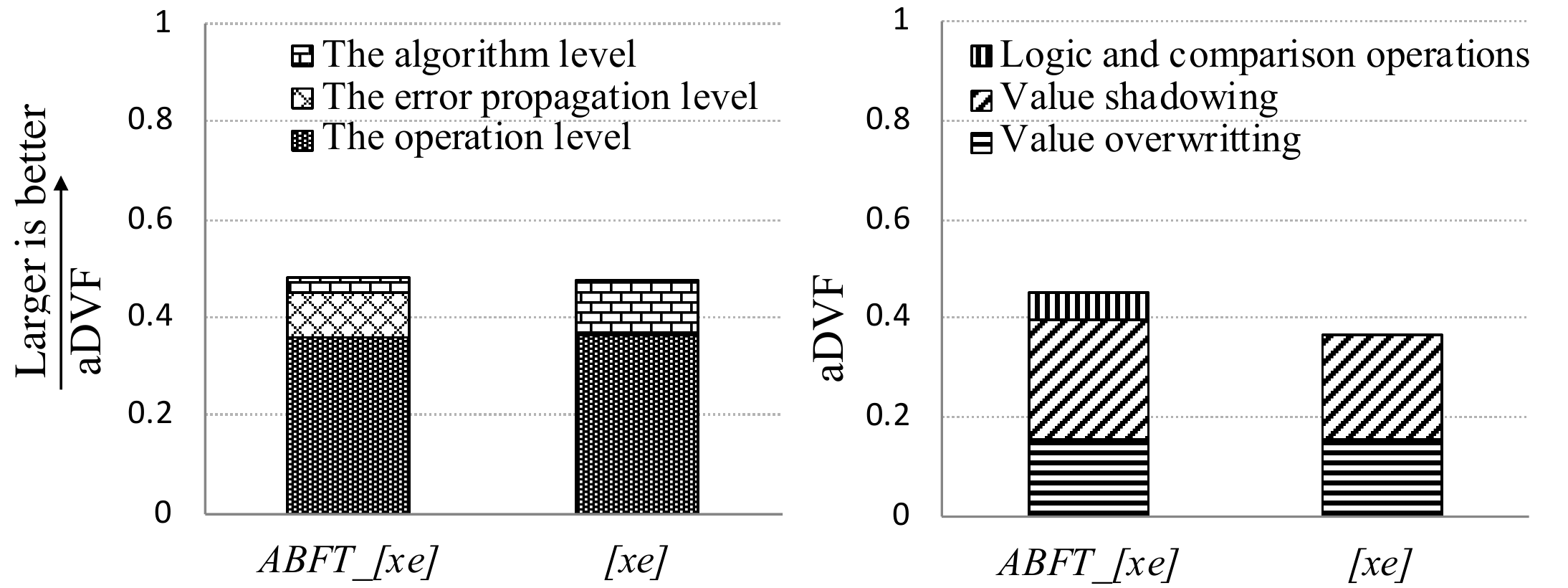}
	\caption{Using aDVF analysis to study the effectiveness of ABFT for a data object $xe$ in PF. [$xe$] has no protection of ABFT; ABFT\_[$xe$] has ABFT taking effect on $xe$.}
	\label{fig:abft_advf_qbox}
    \vspace{-15pt}
\end{figure}

Figure~\ref{fig:abft_advf_qbox} shows the results. The figure reveals that using ABFT does not improve much
application resilience to transient faults on the data object $xe$:
there is only little change to the aDVF value (0.48 vs. 0.475). We find two reasons for it: (1) The operation-level error masking accounts for a large part of error masking, no matter whether we use ABFT or not; (2) Most errors corrected by ABFT are also correctable by PF.
The second reason is demonstrated by the following fact: with ABFT, the number of error masking events increases at the error propagation level but decreases at the algorithm level. 
But in total, the number of error masking events at the both levels with ABFT is almost the same as without ABFT. This case study is a clear demonstration of how powerful the aDVF analysis can direct error tolerance designs.

\section{Discussions}
\label{sec:discussion}
\subsection{Program Optimization by aDVF}
aDVF has many potential usages. We discuss two cases that use aDVF to optimize programs.

\textbf{Code optimization}:
Programmers have been working on code optimization to improve performance and energy efficiency. However, the impact of code optimization on application resilience is often ignored. 
There are cases where optimizing code to improve application resilience is necessary (e.g.,~\cite{hp_sc_outerspace} and~\cite{Maeng:2017:AIE:3152284.3133920}). The code optimization (including common compiler optimization on applications) can change memory access patterns and runtime values of data objects, which in turn impacts error propagation and value shadowing. aDVF and its analysis give programmers a feasible tool to study and compare application resilience (from the perspective of data objects) before and after code optimization. The aDVF analysis is also helpful to pinpoint which part of the application code is vulnerable from the perspective of data objects, and hence demands further optimization.


\textbf{Algorithm choice}:
To solve a specific computation problem, we can have multiple algorithm choices.
For example, to solve the Poisson's equation on a 2D grid, we could use direct method (Cholesky factorization), 
Multigrid, or red-black successive over relaxation. 
Different algorithms have different implications on data distribution, parallelism, and blocking~\cite{pldi09:ansel}.
Which algorithm should be employed depends on users' requirements on performance, energy/power efficiency and resilience.
aDVF and its analysis can help users (especially those users working on HPC) make the algorithm choice from the perspective of application resilience. It would be also interesting to integrate the aDVF analysis with programming language and compiler for algorithm choice, such as PetaBricks~\cite{pldi09:ansel}.

\subsection{Beyond Single-Bit Errors}

MOARD and aDVF calculation are general, meaning that they can be used for analyzing single-bit errors and multi-bit errors. In our study and evaluation, we focus on single-bit errors for two reasons: (1) Multi-bit errors rarely occur in HPC systems, and most of the existing studies on application resilience focus on single-bit errors; (2) Existing work reveals that multi-bit errors can have similar effects as single-bit errors on applications~\cite{DBLP:conf/dsn/SangchooliePK17}.

To use MOARD and aDVF for analyzing multi-bit errors, we need to make the following extension. (1) Define multi-bit error patterns. For example, for two-bit errors, the error pattern could be spatially contiguous; it could also be spatially separated (the spatial separation is four bits, for example). (2) Re-implement the function $f$ (defined in Equation~\ref{eq:dvf}) in MOARD. This indicates that we must re-examine error masking. For the operation-level analysis, the effects of logical and comparison operations and value overshadowing will be different from that for single-bit errors; the effect of value overwriting may be the same as that for single-bit errors. For the error propagation analysis, we can use the same method as for single-bit errors to track error propagation, but the empirical bound of error propagation (i.e., the parameter $k$ in Section~\ref{sec:fault_propagation_analysis}) must be reset using fault injection tests. For the algorithm-level analysis, we use the same fault injection-based method as for single-bit errors,  but the injected errors must follow the defined error pattern.

\subsection{Impact of Input Problems}
The aDVF analysis is input dependent. This means that an application with different input problems may have different aDVF values for a data object. Such input dependence is because of multiple reasons. \textit{First}, the effectiveness of operation-level error masking is input dependent. For example, a bit shifting operation for integers, $x >> y$, can tolerate a single bit error in the least significant bit of $x$ if $y=1$, but can tolerate three single bit errors in the three least significant bits of $x$ if $y=3$. \textit{Second}, different input problems can result in different control flows, which in turn results in different error propagation. \textit{Third}, different input problems can result in the employment of different algorithms. Different algorithms can result in different algorithm-level error masking. 

Because of input dependence nature of the aDVF analysis, we must do the aDVF analysis whenever the application changes its input problem. 
This is a common limitation for many resilience study, including fault injection, AVF~\cite{isca05:mukherjee, micro03:mukherjee}, PVF~\cite{hpca09:sridharan}, DVF~\cite{dvf_sc14} and~\cite{ipdps16:vishnu}. 
However, a static analysis-based method cannot address the limitation because of unresolved branches and data values. Fortunately, MOARD allows a user to easily leverage hardware resource to parallelize the analysis (e.g., deterministic fault injection and trace analysis), making the analysis easy and efficient, even if the user has to repeatedly do the aDVF analysis. Furthermore, leveraging common iterative structures of HPC applications, analyzing a small trace of the application instead of the whole trace is often enough. This makes the repeated aDVF analysis even more feasible. 
Nevertheless, studying the sensitivity of aDVF analysis to input problems is our future work. 

\section{Related Work}
\textbf{Application-level random fault injection.}
Casa et al.~\cite{mg_ics12} 
study the resilience of an algebraic multi-grid solver 
by injecting errors into instruction output based on LLVM. 
Similar work can be found in~\cite{europar14:calhoun, prdc13:sharma}.
Li et al.~\cite{bifit:sc12} build a binary instrumentation-based fault injection tool 
for random fault injection.
Hari et al.~\cite{asplos12:hari, isca14:hari} 
aggressively employ static and dynamic program analyses 
to reduce the number of fault injection tests.
Menon and Mohror~\cite{ppopp18:Menon} apply algorithmic differentiation to 
predict the impact of a SDC on application output to avoid fault injection.
Those research efforts do not sufficiently consider application semantics (e.g., algorithm-level fault tolerance), 
hence provide limited guidance to some application-level fault tolerance mechanisms. 
However, those research efforts can complement our work by accelerating fault injection.
Vishnu et.al~\cite{ipdps16:vishnu} associate data semantics with fault injection to build a machine learning model to predict application errors. However, the data semantics is only introduced at main memory, not registers; also the machine learning model has to be trained and has no accuracy guarantee. Our method does not have the above limitation.

\textbf{Resilience metrics.}
Architectural vulnerability factor (AVF) is a hardware-oriented metric
to quantify the probability of an error in a hardware component resulting in incorrect application outcomes. It was first introduced  in~\cite{isca05:mukherjee, micro03:mukherjee}
and then attracted a series of follow-up work. This includes
statistical modeling techniques to accelerate AVF estimate~\cite{ hpca09:duan},
online AVF estimation~\cite{isca08:Li}, 
Yu et al.~\cite{dvf_sc14} introduce a metric, DVF. DVF 
captures the effects of application and hardware on error tolerance of data objects. 
In contrast to AVF and DVF, aDVF is a highly 
application-oriented metric.
\vspace{-5pt}
\section{Conclusions}
\label{sec:conclusions}
Understanding application resilience (or error tolerance) in the presence of hardware transient faults on data objects is critical to ensure computing integrity and enable efficient application level fault tolerance mechanisms. The traditional methods (such as random fault injection) cannot help because of losing data semantics and insufficient information on how and where errors are tolerated. This paper introduces a fundamentally new method to quantify application resilience to transient faults on data objects. In essence, our method measures error masking events at the application level and associates the events with data objects. We perform a comprehensive classification of error masking events and create a series of techniques to recognize them. We develop an open source tool to quantify application resilience from the perspective of data objects. We hope that our method can make the quantification a common practice. Currently, the deployment of fault tolerance mechanisms is often a problem because of a lack of a method to quantify its effectiveness on protecting data objects. Our work provides a tangible solution to address the problem.

\textbf{Acknowledgements.} This work is partially supported by U.S.   National Science Foundation (CNS-1617967, CCF-1553645 and CCF-1718194) and LLNL subcontract B629135. We thank anonymous reviewers for their valuable feedback.   

\bibliographystyle{IEEEtran}
\bibliography{li,li_sc16_resilience_modeling}  

\end{document}